# Accurate Stress Assessment based on functional Near Infrared Spectroscopy using Deep Learning Approach

Mahya Mirbagheri
School of Electrical and Computer Engineering
College of Engineering, University of Tehran
Tehran, Iran
Mt.mirbagheri@ut.ac.ir

Ata Jodeiri
School of Electrical & Computer Engineering
College of Engineering, University of Tehran
Tehran, Iran
ata.jodeiri@ut.ac.ir

Naser Hakimi
School of Electrical and Computer Engineering
College of Engineering, University of Tehran
Tehran, Iran
naserhakimi@ut.ac.ir

Vahid Zakeri
Simon Fraser University
Vancouver, BC, Canada
vza2@sfu.ca

Seyed Kamaledin Setarehdan
School of Electrical & Computer Engineering
College of Engineering, University of Tehran
Tehran, Iran
ksetareh@ut.ac.ir

*Abstract*—Stress is known as one of the major factors threatening human health. A large number of studies have been performed in order to either assess or relieve stress by analyzing the brain and heart-related signals. In this study, signals produced by functional Near-Infrared Spectroscopy (fNIRS) of the brain recorded from 10 healthy volunteers are employed to assess the stress induced by the Montreal Imaging Stress Task by means of a deep learning system. The proposed deep learning system consists of two main parts: First, the one-dimensional convolutional neural network is employed to build informative feature maps. Then, a stack of deep fully connected layers is used to predict the stress existence probability. Experiment results showed that the trained fNIRS model performs stress classification by achieving 88.52 ± 0.77% accuracy. Employment of the proposed deep learning system trained on the fNIRS measurements leads to higher stress classification accuracy than the existing methods proposed in fNIRS studies in which the same experimental procedure has been employed. The proposed method suggests better stability with lower variation in prediction. Furthermore, its low computational cost opens up the possibility to be applied in real-time stress assessment.

## I. INTRODUCTION

Stress is one of the major factors leading to chronic disorders [1], [2]. It is a common physical response to the environment making people feel uncomforted, challenged or threatened [3]. Stress disturbs imagination, problem-solving, decision making, working memory and other related prefrontal cortex (PFC) activities [4], [5]. Therefore, precautionary measures to assess and then relieve stress are vitally important for both individual health and the welfare of society at a broader level.

Appropriate standardized protocols for studies of stress are required to induce stress in a reliable and credible way. Montreal Imaging Stress Task (MIST) is one of the popular psychological stressor protocols employed in the stress assessment studies [6]. In the MIST, mental arithmetic calculations are performed during a stressful condition (in a limited time) while the participants' response is assessed [7].

To study the development of stress, continuous measurements are required to study the mental state throughout consecutive and stressful tasks over a long period [8], [9]. It has been shown that stress can be assessed by brain-related physiological variables including EEG (Electroencephalography) and functional near-infrared spectroscopy (fNIRS) [10]–[15]. In the past decade, mental stress and workload assessments [16], [17] have been addressed by employing the fNIRS as a measure for quantification, showing an increased concentration of oxyhemoglobin (HbO2) and decreased concentration of deoxyhemoglobin (HHb) in the prefrontal.

For stress detection utilizing physiological signals, an intelligent system based on machine learning approach is needed. Several studies have shown the superiority of deep learning as the main sub-fields of machine learning in data analysis tasks [18], [19]. The advantage of deep learning is caused by its ability to automatically learn high-level, layered, hierarchical abstractions from data and mapping the extracted features to desired labels by end-to-end training [20]. For automatic feature extraction, a type of deep artificial neural networks named convolutional neural networks (CNN) is widely used in the field of computer vision [21]. CNNs have been applied to many tasks, including image classification [22], object detection [23], and semantic segmentation [24]. Although CNNs was introduced for images, several studies have shown the CNN's ability in extracting deep features from one-dimensional data such as speech signals [25] and vector of raw input data [26].

Motivated by promising results of deep networks in various fields [27]–[31], in this study we investigated its applicability in stress prediction by means of fNIRS which to the best our knowledge has not been done. The stress protocol conducted in this study is a modified version of the MIST [7], designed and used to induce a higher level of mental stress. An fNIRS dataset is made by extracting features from both time and frequency domains of fNIRS signals. This study proposes a deep learning based algorithm that can figure out specific abstract patterns in each data to determine the distinctiveness of each case and is able to map the input data directly to their corresponding label, determining whether it is stress related or not.

## II. MATERIALS AND METHODS

### A. Experiment Procedure

Ten healthy, right-handed, male adults (aged 25.3 ± 2.6) participated in this experiment. It should be mentioned that none of them had psychological and neurological diseases and did not take any special medication. Before experimenting, they were informed about the experiment and gave written consent. Signals were registered at the Iranian National Brain Mapping Lab. The laboratory had an environment with a good air-condition without any environmental stress and noises. Participants were seated in a comfortable chair, located approximately 1.5 meters away from the monitor while they were asked to avoid shaking during the experiment as much as possible (Figure 1). The experiment has an ethics code of IR.IUMS.REC.1396.810194120 from the Iranian University of

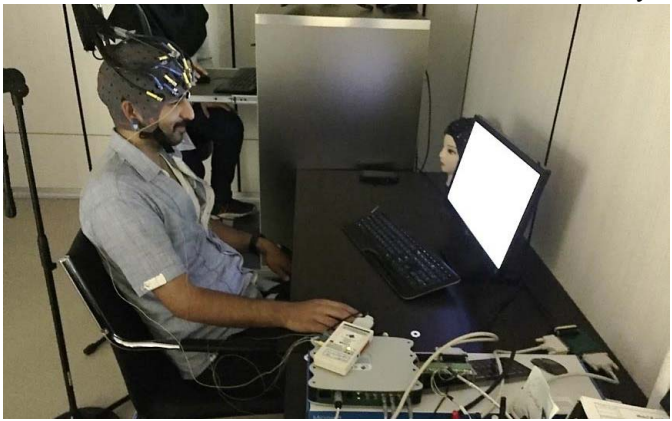

Figure 1. The Experiment Procedure for participant conducting the task. The experiment was performed at the Iranian National Brain Mapping Lab.

Medical Sciences.

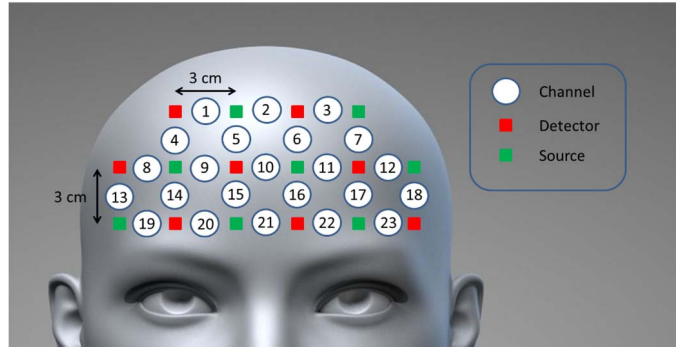

Figure 2. The location of the fNIRS detectors and sources, and the layout of the 23 fNIRS channels. The source-detector distance was considered 3 cm.

### B. Data Acquisition Protocol

Brain activity was recorded by placing 23 fNIRS channels at the PFC region. The location of transmitters, receivers, and the layout of the channels are illustrated in Figure 2. The fNIRS system employed in this study was OxyMonfNIRS (Artinis Medical System, Netherland), which is equipped with light wavelengths of 845 and 762 nm, sampling rate of 10 Hz, and transmitter-receiver distance of 3 cm.

### C. Task Sequence

The experiment was designed based on the MIST protocol [7] to induce mental stress. Several modifications have been made to reform the task according to the studies of [32], [33]. The task includes four levels of Explanation, Training, Control, and Stress. No signal is taken in Explanation and Training levels, while signals which were mentioned in the Data Acquisition Protocol section are recorded during Control and Stress levels.

Concerning the task sequence, the signals of each participant are recorded in approximately 10 minutes including 5 minutes for each of the control and Stress levels. In this regard, Control and Stress levels have five two-part blocks, the first parts of which are related to 20 seconds of rest, and the second parts are related to 30 seconds of solving the mathematical questions.

### D. Signal Analysis

In the first step of this study, fNIRS signals are acquired through a protocol which was discussed in section of Data Acquisition Protocol. Secondly, fNIRS signals are preprocessed by conducting two methods of ICA Denoising and Band Pass Filtering before the stress classification process. Finally, the preprocessed fNIRS signals are employed to classify mental stress based on a deep learning method.

*1) fNIRS Signal Preprocessing*

In this section, ICA is conducted to suppress noises of fNIRS signal related to physiological oscillations including heart, and the other unknown artifacts related to the cardiac function according to [34], [35]. On purpose, a cardiac wave is firstly simulated representing the heart noise in the fNIRS signal based on the method described in [36]. The cardiac wave is constructed by integrating the Gaussian waves with constant amplitude and standard deviation of 0.05s which are centered in the times heart beats are detected based on the method described in [33]. Next, the independent components of fNIRS signals are computed and compared with the simulated heart wave. Finally, 20% of components have the lowest correlation with the simulated wave are reconstructed. The optimum percentage of the selected components has been obtained by doing try and error.

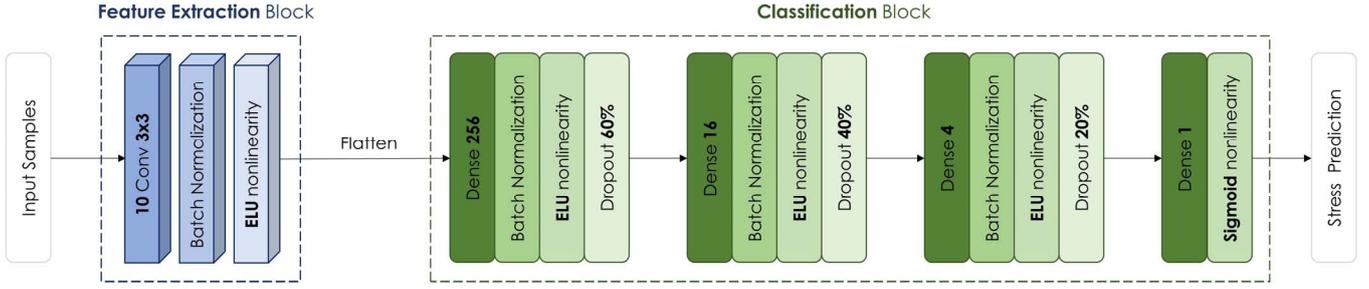

Figure 3. Network architecture consists of feature extraction block and classification block.

In addition to ICA denoising, a band-pass filter of [0.001-0.14] Hz is applied to the fNIRS signal before classification and comparison processes in order to eliminate the unrelated frequency components to brain hemodynamic response [37], [38].

*2) Stress Classification*

To describe the feature extraction process, in addition to four features in the time domain including $m_t$, $sd_t$, $Sk_t$, and $Ku_t$, four features in the frequency domain including $m_f$, $sd_f$, $Sk_f$, and $Ku_f$ have also been extracted from fNIRS signals. All in all, eight different features are considered for fNIRS signals. The selected features have been described in studies [33], [39].

As mentioned in Task Sequence section, Control and Stress levels have five two-part periods including 20 seconds Rest, and 30 seconds mathematical questions. In this study, a classification procedure is conducted which is based on the study [32]. In this procedure, a classification of the 30-second parts between Control and Stress levels is pursued.

To form an fNIRS dataset, ten 30-second parts are separated from each recorded fNIRS signal. Then each part is divided into three 10-second sections, and the considered eight features for fNIRS representation are extracted from each of these three sections. This process is applied to both oxyhemoglobin and total hemoglobin signals of fNIRS channels, which is the best combination based on our results. The constructed fNIRS datasets are fed into a deep neural network architecture which is explicitly explained in the following and illustrated in Figure 3.

The network consists of two main blocks: feature extraction and classification blocks which are connected to each other by flattening layer. For automatic extraction of feature maps from raw input samples, a one-dimensional convolutional neural network with ten 3x1 kernel size, Batch Normalization and ELU non-linearity are used. The classification block contains five blocks in which three first blocks are made from a Dense (fully-connected) layer, Batch Normalization, ELU non-linearity, and Dropout layer. In the last block, we used one neuron layer and Sigmoid function. The numbers of activation neurons of Dense layers are 256, 16, 4, and 1 respectively. In order to prevent over-fitting problem due to the relatively large number of trainable parameters, three 60%, 40% and 20% drop-out layers are added after the ELU nonlinearities in three first blocks respectively.

The proposed model is implemented using Tensorflow library [40] and is optimized by minimizing Adam algorithm with 0.0002 learning rate. This task is treated as binary classification problems, where each output neuron decides if a sample belongs to a class or not, so binary cross-entropy is used for loss function which Adam algorithm [41] tries to decrease it through the epochs. At each epoch, a mini-batch size of 200 samples is randomly selected from the training images and is fed into the network. It is notable that for training the network with fNIRS data, the batch size is set to 20. The network will converge in 500 epochs, and neither data augmentation nor pre-train weights is used during training.

In order to show that the results do not depend on the choice of test dataset, the five-fold cross-validation strategy has been employed based on the articles [33], [38], [42] which use cross validation strategy. In five-fold cross-validation, the original sample was randomly partitioned into five equal-sized subsamples. In each experiment, a single subsample was retained for testing the model, and the remaining four subsamples were used for training the networks. The cross-validation process was then repeated five times, with each of the five subsamples used exactly once as the test dataset.

III. RESULT

An example of the filtered fNIRS signals averaged between channels and recorded from the ninth participant is illustrated in Figure 4.

The introduced ICA denoising and band-pass filter in fNIRS Signal Preprocessing section have been used to remove artifacts and noises irrelevant to the brain hemodynamic response. The first five green parts and the second five orange parts are related to the Control and Stress levels, respectively. After entering to the Stress level, a significant difference in the signal variation is visually observable.

Based on the 5-fold cross-validation strategy defined in the Stress Classification section, Table I reports the accuracy of network prediction in each fold in addition to the mean and standard deviation (STD) of the accuracy over five-fold strategy. In particular, the prediction accuracy obtained using fNIRS is 88.52 ± 0.77%.

TABLE I
NETWORK ACCURACY ON DIFFERENT TRAIN AND TEST DATASETS CORRESPONDING TO 5-FOLD CROSS-VALIDATION

| Signal | Accuracy (%) | | | | | Mean ± STD (%) |
|---|---|---|---|---|---|---|
| | Fold#1 | Fold#2 | Fold#3 | Fold#4 | Fold#5 | 5-Fold |
| fNIRS | 89.27 | 89.04 | 88.52 | 88.72 | 87.07 | 88.52 ± 0.77 |

In the Stress Classification section, it was mentioned that each 30-second desired part of the signal is divided into three 10-second sections to form the dataset, and the considered features were extracted from each of these three sections. To identify the information richness level of these three time frames, stress classification was separately performed by using each of these time frames. The related results of the classification are reported in Table II. It is observed that the second and third time frames of fNIRS signals respectively contain much more levels of information richness than other corresponding time frames, where 85.14% of mean accuracy was obtained by using features extracted from the mentioned best time frames of fNIRS signals.

TABLE II
STRESS DETECTION MEAN ACCURACY USING DATASET INCLUDING FNIRS CONSIDERING DIFFERENT TIME FRAMES AND FEATURE SETS

| Dataset | fNIRS (%) |
|---|---|
| Time Frame #1 | 71.11 |
| Time Frame #2 | 78.32 |
| Time Frame #3 | 85.14 |
| First Feature Set | 86.10 |
| Second Feature Set | 80.74 |
| All | 88.52 |

In this study, in addition to features from the time domain of the signals, some features in the frequency domain of the signals were also extracted to represent the variations of the signals as better and efficient as possible. To show the effectiveness of these time and frequency domain features, stress classification was separately performed by using each of the time and frequency feature sets. The related results of classification are reported in Table II, indicating that the time domain features (First Feature Set) can better modulate the stress related variations rather than the frequency domain feature set (Second Feature Set), or in other words, the stress-related alterations are significantly made in the time domain of the signal. For example, utilizing the time domain features of the fNIRS signal could provide the mean stress classification accuracy of 86.10% compared with 80.74% of accuracy by the employment of the frequency domain features.

However, incorporation of time domain features with the frequency ones could provide a better stress classification accuracy. So that by using both time and frequency domain features of fNIRS signals, a mean stress classification accuracy of 88.52% was obtained.

The size of the training data is one of the determinant factors in the results obtained from the deep networks. Naturally, with increasing the number of training samples, the network performance is expected to be improved. On the other hand, a model is better which could discover the patterns needed for decision making using minimum possible data. In this experiment, the number of training data has gradually decreased, while the test data size has been kept constant to see the reaction of three models. The mean accuracy of the trained model by fNIRS over five folds is reported in Table III.

By comparing the performance of the models trained on different data sizes, we conclude that fNIRS model has high stability against a reduction in input samples. For example, with 20% drop from the whole dataset, the accuracy of fNIRS model only was decreased by 3%. The results of the table indicate that the trained model on fNIRS dataset is robust, in which the trained model with only 20% fraction of the data achieved 78.58% accuracy.

TABLE III
STRESS DETECTION MEAN ACCURACY USING DATASET INCLUDING FNIRS CONSIDERING THE VARIOUS FRACTION OF TRAIN DATASET

| Fraction of Train Dataset (%) | fNIRS (%) |
|---|---|
| 100 | 88.52 |
| 80 | 85.57 |
| 60 | 82.33 |
| 40 | 81.54 |
| 20 | 78.58 |

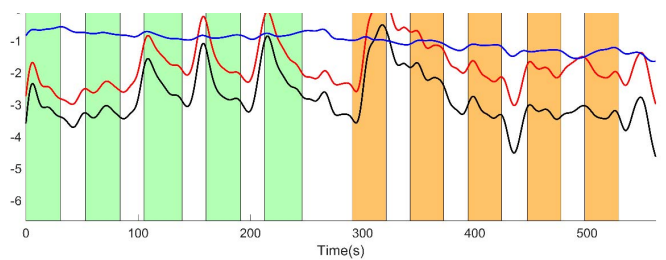

Figure 4. An example of the averaged fNIRS signals recorded from the ninth participant during the MIST. The first five green parts are corresponding to the Control level and the second five orange parts are related to the Stress level. The Black, Red, and Blue curves are related to the total, oxyhemoglobin, and deoxyhemoglobin signals, respectively.

IV. DISCUSSION

This study investigated the applicability of deep learning in stress prediction by using fNIRS signal, a novel idea presented for the first time. In this regard, we assessed mental states of 10 healthy right-handed male subjects under the induction of a stressful condition by MIST. Before performing the stress assessment, fNIRS signals were preprocessed. Next, extracting efficient features from both time and frequency domains of fNIRS signals, a deep learning based algorithm was applied to determine the distinctiveness of each case and able to map the input data directly to their corresponding label. This algorithm was applied in order to propose a reasonably comprehensive signal, feature set, and deeply learned system to be sufficiently successful in this regard.

Our results show that the fNIRS provides an effective means for prediction of the stress during the MIST in the sense of providing higher classification accuracy. Using the features extracted from fNIRS signal, stress was classified employing intelligence model based on deep neural network architecture with the mean accuracy of 88.52%. When it comes to STD, it can be observed that the fNIRS model could provide an STD accuracy of 0.77%. Therefore, the proposed model trained on fNIRS not only provides a viable system for stress detection but also suggests better stability in terms of accuracy of the network's prediction which ensures us the system's performance is not related to the choice of train and test dataset.

In this study, a highly sufficient stress predictor is proposed by employing fNIRS signal which is real-time and non-invasively recorded. By broadcasting the artificial neural networks as an exclusive method for classification, a new deep learning model is introduced in this study in order to bring accurate and reliable prediction beside to low computational cost, making it totally suitable for wearable and real-time applications [43]. Employment of the proposed deeply learned system with the extraction of efficient features from the fNIRS signals lead to a mean classification accuracy of 88.52% which is higher than the accuracies reported in [32], [33], [38], [42], in which the same stress protocol and the same signal were employed. However, one of the limitations of our study is that a less number of subjects were investigated compared with their study. It will be interesting to investigate the applications of the proposed system in such similar research fields of attention, mental workload, mental diseases, and sleep [43]–[50].

In this study, to determine the information richness level of the three 10-second time frames considered for feature extraction, stress classification was separately performed by using the extracted features from each of these time frames. It was observed that the third time frame of fNIRS signals is informative than other corresponding time frames. It confirms the natural delay of the hemodynamic response in representing the brain related changes induced by mental stress in the fNIRS signal.

As two types of features have been extracted to convey the stress-related information as efficient as possible; frequency and time domain features, to show the effectiveness of these features, stress classification was separately performed by using each of the time and frequency feature sets. The results indicate that the time domain features can better modulate the stress-related information compared with the frequency domain feature set so that by employing four features from the time domain of the fNIRS signal, we could predict the stress induced by MIST with a remarkable accuracy of 86.10%, which is applicable for real-time applications where calculation time is restricted. However, incorporation of frequency domain features into the time domain ones could provide a better stress classification accuracy.

Regarding the model's mean accuracy in a different fraction of dataset, it is concluded that the fNIRS model performs robust. In particular, when 80% of the training data were excluded, the mean accuracy of fNIRS models was only decreased almost by 10% to 78.58%. In other words, fNIRS could provide an effective representation of stress-related patterns and performance.

This study proposes an approach to measure a signal arising from brain which can be employed for other applications besides stress assessment in which brain function is altered including biofeedback [51], BCI [52]–[55], infant studies [56]–[59]. The present study was limited to the PFC region while such applications might also require recording the brain activities of other brain regions. So investigating the neuronal activities of those regions can be pursued as a future study.

V. CONCLUSION

In this study, fNIRS signals were employed to assess the stress induced by the Montreal Imaging Stress Task using a deep learning system. On purpose, mental states of 10 healthy volunteers were assessed by recording and then preprocessing the fNIRS signals. In our proposed method, feature maps were automatically extracted utilizing the convolutional network, and then a stack of fully connected layers classified the feature maps to determine the probability of stress in input data. With several experiments, we showed the high performance of our proposed method trained on fNIRS dataset in terms of prediction accuracy, model stability and fast performance applicable for real-time stress assessment.

ACKNOWLEDGMENT

This project was partially supported by NBIC council (Nanotechnology, Biotechnology, Information Technology & Cognitive Sciences), NBML (National Brain Mapping Lab), and CSTC (Cognitive Sciences and Technologies Council) of Iran. The authors would like to thank all the people who participated in the study, including subjects and students that collaborated.